\documentstyle[aps,preprint,prc,axodraw]{revtex}

\begin{document}

\tighten
\draft
%%\twocolumn
\preprint{
\vbox{
\hbox{{\tt INR report 0967/97}}
\hbox{{\tt nucl-th/9801039}}
}}

\title{
Nuclear Effects In The $F_3$ Structure Function  \protect\\
For Finite And Asymptotic $Q^2$
}

\author{S. A. Kulagin}

\address{
        Institute for Nuclear Research,
        Russian Academy of Sciences,
        Moscow, Russia
}

\maketitle

\begin{abstract}

\noindent
We study nuclear effects in the structure function $F_3$ which
describes
the parity violating part of the charged-current
neuitrino nucleon deep inelastic scattering.
Starting from a covariant approach we derive a factorized expression
for the nuclear structure function in terms of nuclear spectral
function and off-shell nucleon structure functions valid for
arbitrary momentum transfer $Q$ and in the limit of weak nuclear
binding, i.e. when a nucleus can be treated as a non-relativistic
system.
We develop a systematic expansion of nuclear structure functions
in terms of $Q^{-2}$ series caused by nuclear effects
(``nuclear twist" series).
Basing on this expansion we calculate nuclear corrections to
the Gross-Llewellyn-Smith sum rule as well as to higher moments
of $F_3$.
We show that corrections to the GLS sum rule due to nuclear effects
cancel out in the Bjorken limit and calculate the corresponding
$Q^{-2}$ correction.
Special attention is payed to the discussion of the off-shell effects
in the structure functions.  A sizable impact of these effects both
on $Q^2$- and $x$-dependence of nuclear structure functions is found.

\vskip 1cm
%\noindent
%Keywords:
%deep inelastic scattering,
%neutrino, charge current,
%impulse approximation,
%EMC-effect,
%twist expansion,
%off-shell amplitudes.

\end{abstract}

\newpage
%%%%%%%%%%%%%%%%%%%%%%%%%%%%%%%%%%%%%%%%%%%%%%%%%%%%%%%%%%%%%%%%%%%%%%%
\section{Introduction}

Experiments on deep-inelastic scattering (DIS) of charged leptons and
neutrino remains an important source of information about nucleon and
nuclear structure.
Let us note in this respect that, though the primary aim of these
experiments is to explore the structure of hadrons, DIS data is
collected usually for nuclear targets.
Since the discovery of nontrivial
nuclear effects in DIS with muon and electron beams (the EMC effect),
the studies of nuclear structure functions
is an important topic of itself
(for a recent reviews of experimental situation and theoretical
approaches to nuclear effects in DIS see \cite{Arneodo94,GST95}).

In the literature nuclear effects have been discussed in detail for
the spin independent structure functions $F_{1,2}$ as well as for
spin structure functions $g_{1,2}$ which are measured in the
charged lepton DIS. However up to now only a little attention
was payed to nuclear effects in neutrino DIS
\cite{KM93,KKM95,SiTo95} and
especially to the structure function $F_3$ which does not have an
analog in the DIS with the charged leptons.
Perhaps such a lack of interest can be partially explained
by the lack of precise neutrino data for different targets,
that would allow to separate nuclear effects.
We recall also that in the parton model
$F_3$ is related to the valence quark distribution and at large
$x>0.3$, where the sea part is small, $xF_3$ is practically equal
to $F_2$. It is expected therefore that nuclear modifications of
$xF_3$ is similar in this region to those of $F_2$,
which has been studied in more details.
It should be stressed however that
this similarity between $F_2$ and $xF_3$
is only approximately valid in a certain kinematical region
and fails at small $x$, where the sea part of
the parton distributions is important, as well as
when one studies effects due to finite $Q^2$.
We note also that a precise neutrino data is now available
for the iron target \cite{CCFR:96}
(for a review on neutrino data see Ref.\cite{CSB97}),
and in order to perform a QCD analysis \cite{KKPS96} of the data
one should have a reliable method to handle nuclear effects
in the neutrino DIS at finite $Q^2$.

These arguments motivate us to study the neutrino and anti-neutrino DIS
from nuclear targets, that we do in the present paper.
Our primary aim here is to study nuclear effects in
the structure function $F_3$ within approach applicable
at finite $Q^2$ and develop a reliable method to
extract $F_3$ of the isolated proton and neutron from nuclear data.
For this purpose we re-examine and extend to finite $Q^2$
the impulse approximation formalism
for nuclear structure functions, developed previously
for large $Q^2$ for the structure functions $F_{1,2}$
\cite{Ku89,KPW94},
as well as for the spin structure functions $g_{1,2}$
\cite{KMPW95}.

The paper is organized as follows.
In Secs.\ref{framework},\ref{nuclearSF} we present a general
covariant framework in which nuclear structure functions are
expressed in terms of nucleon propagator in the nucleus and the
off-shell nucleon hadronic tensor.
Sec.\ref{nr-limit} describes how the relativistic expressions for
nuclear structure functions can be reduced
in the limit of weak nuclear binding by making a
non-relativistic expansion of nuclear matrix elements.
In Sec.\ref{sf_nr} we analyze the resulting generalized convolution
equation for the nuclear structure function and develop its
systematic expansion in nuclear $Q^{-2}$ series.
In Sec.\ref{offshell} we discuss off-shell
effects in the nucleon structure function.
Nuclear corrections to the Gross-Llewellyn-Smith sum rule as well as
to higher moments of $F_3$ are calculated in Sec.\ref{sumrule}.
Sec.\ref{D-expan} deals with an approximate expression for the
nuclear structure function which arises if one treats the effects of
Fermi-motion and nuclear binding perturbatively.
We summarize and make concluding remarks in Sec.\ref{finita}.

% II %%%%%%%%%%%%%%%%%%%%%%%%%%%%%%%%%%%%%%%%%%%%%%%%%%%%%%%%%%%%%%%%
\section{Hadronic Tensor and Structure Functions}
\label{framework}

We consider the neutrino (anti-neutrino) charged-current interaction.
Inclusive inelastic scattering from hadrons is described by the
hadronic tensor
\begin{eqnarray}
W_{\mu \nu}^{\pm}(P,q,S)
&=& \frac{1}{8\pi} \int d^4z\, e^{iq\cdot z}
\left\langle P,S \left| \left[ J_{\mu}^\pm(z), J_{\nu}^\mp(0)
			\right]
		 \right|P,S
\right\rangle,
\label{wmunu}
\end{eqnarray}
where $J_\mu^\pm$ is the charged-current corresponding to the exchange of
$W^\pm$-boson, $P$ and $q$ are the four-momenta of the target and
virtual $W$-boson, respectively, and the vector $S$ is the target
polarization (spin) vector.  In what follows we discuss scattering of
unpolarized neutrino beams from unpolarized targets and will drop
an explicit polarization dependence of the hadronic tensor.

It is well known (see, e.g., \cite{IoKhLi84})
that unpolarized neutrino nucleon deep-inelastic
scattering is described in general by five structure functions. Two
of those are similar to the structure functions $F_1$ and
$F_2$ in charged lepton DIS.
The structure function $F_3$ arises due to
vector--axial current interference in Eq.(\ref{wmunu})
and determines the $\nu$-$\bar\nu$ cross-section
asymmetry.  The left two structure functions, $F_4$ and $F_5$, are
due to nonconservation of the axial current. These structure
functions are usually not discussed in the DIS regime,
because their contribution to
the cross-section is suppressed by a tiny ratio of the lepton mass to
the neutrino beam energy.  Apart of this argument, $F_4$ and $F_5$
have some ``internal smallness" due to PCAC \cite{IoKhLi84}.

In the present paper we focus on the discussion of
the vector-axial current interference term which contribution
to the hadronic tensor (\ref{wmunu}) reads:
\footnote{
We use the following normalization of the states throughout of the paper
$\left\langle p|p' \right\rangle=(2\pi)^3 2p_0\delta({\bf p}-{\bf p'})$.
In this normalization the hadronic tensor (\ref{wmunu}) is dimensionless.}
\begin{eqnarray} \label{W3}
W_{\mu\nu}(P,q)
&=& i\,\epsilon_{\mu\nu\alpha\beta}\,P^\alpha q^\beta
     \frac{F_3(x,Q^2)}{2\,P\!\cdot\!q},
\end{eqnarray}
where the four-momentum transfer squared $Q^2=-q^2$ and
the Bjorken variable $x=Q^2/2P\cdot q$ are the standard
DIS variables.

%....................................................................
%%\subsection{Nucleon Tensor and Structure Functions}

In our covariant analysis of neutrino-nucleus scattering
it will be useful to work with the
off-shell nucleon tensor $\widehat {\cal W}_{\mu\nu}$, which is
defined through the imaginary part of the forward $W$-boson
scattering amplitude from an off-shell nucleon.
In terms of $\widehat {\cal W}_{\mu\nu}$,
the hadronic tensor of a ``physical" on-mass-shell nucleon
averaged over nucleon polarizations is:
\begin{eqnarray}\label{What}
W_{\mu\nu}^N(p,q)
&=& \frac12\text{Tr} \left[
(\overlay{\slash}p+M)
             \widehat {\cal W}_{\mu\nu}(p,q) \right],
\end{eqnarray}
were $p$ is the nucleon momentum, $M$ is the nucleon mass and $p^2 = M^2$.

In off-shell region the Lorentz tensor
structure of $\widehat{{\cal W}}_{\mu\nu}$
is more complicated than the corresponding structure
for the on-shell nucleon.
In order to establish the tensor structure
of $\widehat{{\cal W}}_{\mu\nu}$
we expand the latter
in terms of a complete set of Dirac matrices,
$ \left\{ I, \gamma^\alpha, \sigma^{\alpha\beta},
	  \gamma^\alpha\gamma_5, \gamma_5
\right\}$.
The various coefficients in this expansion
must be constructed from the vectors $p$ and $q$,
and from the tensors $g_{\alpha\beta}$ and
$\epsilon_{\mu\nu\alpha\beta}$.
We consider here the anti-symmetric part of the tensor
$\widehat {\cal W}_{\mu\nu}$ and
separate only those terms  which are
even under time-reversal and odd under parity transformations,
since only such terms can contribute to $F_3$.
If we keep only current conserving terms then the most general
form of the antisymmetric tensor $\widehat {\cal W}_{\mu\nu}(p,q)$
which obeys to these requirements can be written as follows,%
\footnote{
The symmetric part of the tensor $\widehat {\cal W}_{\mu\nu}$,
which is relevant for $F_1$ and $F_2$ structure functions,
have been discussed in detail in Ref.\cite{MeScTh94}.
The $T$-even and $P$-even terms
in the antisymmetric part of
$\widehat {\cal W}_{\mu\nu}$, which determine
the spin structure functions $g_1$ and $g_2$,
have been constructed in Ref.\cite{KMPW95}.
}
\begin{eqnarray}\label{Whatdef}
\widehat {\cal W}_{\mu\nu}(p,q)
&=&
i\,\epsilon_{\mu\nu\alpha\beta}\,q^{\alpha}\,
\widehat{\cal G}^{\beta}(p,q)/(4\,p\cdot q),\\
\widehat{\cal G}^{\beta}(p,q)
&=& \left(
         \frac{f^{(0)}}{M}
        + \frac{f^{(p)}}{M^2}\overlay{\slash}p
         + \frac{f^{(q)}}{p\cdot q}\overlay{\slash}q
    \right) p^{\beta}
 + f^{(\gamma)} \gamma^{\beta},
\label{DirSt}
\end{eqnarray}
where the coefficient functions $f^{(i)}$ are
scalar, dimensionless and real functions of
$q^2$, $p\cdot q$ and $p^2$.

We find therefore that an off-shell nucleon is characterized by {\em four}
independent structure functions which correspond to $T$-even and
$P$-odd terms in the hadronic tensor.  On the mass-shell $p^2=M^2$,
because of the Dirac equation, only one out of the four structure
functions in Eq.(\ref{DirSt}) is independent, and Eq.(\ref{Whatdef})
reduces to Eq.(\ref{W3}).  Substituting
Eqs.(\ref{Whatdef},\ref{DirSt}) to Eq.(\ref{What}) we find:
\begin{eqnarray}
\label{F3on}
F_3 &=& f^{(0)} + f^{(p)} + f^{(q)} + f^{(\gamma)},
\end{eqnarray}
with the functions $f^{(i)}$ evaluated here at $p^2=M^2$.

We conclude this section by noting that the nucleon structure function
$F_3$ splits into four structure functions $f^{(i)}$ in off-shell
region. We discuss the concequences of this observation for nuclear
scattering in the next section.

%....................................................................
\section{Nuclear Structure Functions}
\label{nuclearSF}

\begin{figure}[htb]
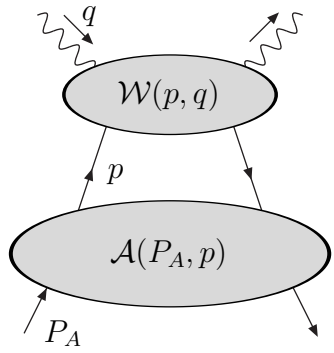

\begin{center}
\vskip 2cm
\input fig1.pic
\end{center}
\vskip 2cm
\caption{Deep-inelastic scattering from a nucleus
in the impulse approximation}
\label{fig1}
\end{figure}

Discussion of nuclear effects in DIS is usually done separately for
small and large Bjorken $x$.
The corresponding scale for this separation comes from the
comparision of a characteristic time for DIS, $1/Mx$ (see, e.g.,
Ref.\cite{IoKhLi84}), and an average distance between bound
nucleons which is about $1\,$Fm for heavy nuclei.
At large $x>0.2$ the characteristic DIS time is smaller
than average internucleon distance.
This observation motivates the use of the impulse
approximation in this region, i.e. the nuclear Compton amplitude is
approximated by the uncoherent sum of the scattering amplitudes from
bound nucleons, Fig.\ref{fig1}.
Major nuclear effects found in this region are due to
Fermi-motion
\cite{fm,binding,Ku89,KPW94,KMPW95}
nuclear binding \cite{binding,Ku89,KPW94,KMPW95},
and off-shell modification of nucleon structure functions
\cite{KPW94,KMPW95,MeScTh94}
(for more discussion and references see review paper \cite{GST95}).

At small $x$ DIS proceeds mainly via diffractive mechanism, i.e.
virtual photon/boson first converts into a $q\bar q$ fluctuation
which subsequently interacts with the target.  Since an average life
time of the $q\bar q$ fluctuation at small $x$ is large as compared
with an average distance between bound nucleons, coherent multiple
scattering effects are important.  It is known that for the structure
function $F_2$ this mechanism leads to nuclear shadowing effect (for
a review of models for nuclear shadowing see, e.g.,
\cite{Arneodo94}).  However a little is known about shadowing effect
in the structure function $F_3$\cite{KKM95}.

Other possible nuclear effects which go beyond the impulse
approximation are corrections due to meson exchange currents (MEC)
and final state interactions (FSI).

While MEC give a sizable correction to the structure function $F_2$ at
small $x<0.2$ (see, e.g., Refs.\cite{mec,Ku89} and a review \cite{GST95}),
one may argue that for
$F_3$ the corresponding correction is less important.  Indeed the structure
function $F_3$ in neutrino scattering in the quark-parton model reads
$F_3^{\nu}(x)=d(x)-\bar u(x)+s(x)-\bar c(x)$.  For light mesons, such
as $\pi$- and $\rho$-mesons, the contribution from $u$- and $d$-quarks
cancels out because of the isospin symmetry. $F_3$ therefore vanishes
for light mesons
if one neglects a contribution due to $s-\bar c$ (similarly $F_3$
vanishes for anti-neutrino scattering up to $\bar s-c$ contribution).
By this reason one can neglect MEC correction to nuclear $F_3$.
\footnote{%
We comment however that MEC may cause a finite contribution
to nuclear $F_3$ due to $s$-quark content of the mesons.
Special care has to be taken of the strange mesons contribution.
}

It is commonly believed that the FSI in DIS is a higher twist effect,
i.e. appears at least in $Q^{-2}$ order.  It has been indeed found
within framework of the collinear expansion \cite{EFP83}, that the
Feynman diagrams describing the FSI of the struck quark with the rest
of the target are suppressed at high $Q^2$.  Similar conclusion is
drawn within Bethe-Solpeter formalism in Ref.\cite{Gur95}.
Cancellation of the FSI in the Bjorken limit is also found
in the studies of inelastic scattering in non-relativistic
\cite{IoKhLi84,fsi1}
as well as in relativistic
\cite{fsi2} two-body models
even with a ``confining interaction" such as oscillator
potential.
These observations might be considered as an argument in favor
of the ``common wisdom'' that cancellation of FSI effects
in total cross-sections
is due to unitarity and closure of the final states.
However the situation is less clear in the case of DIS from nuclei,
where the struck quark may scatter softly from bound nucleons while
before hadronization.
This effect may lead to a finite FSI
correction to the nuclear structure functions
even in the Bjorken limit \cite{Kul_q90}.
Phenomenological arguments in favor of the suppression of the FSI
in nuclear DIS are recently given in Ref.\cite{STZ98}.
Finally we comment that the systematic treatment of the FSI in DIS
from nuclei is still lacking and
remains a challenging problem, which is however out of the scope of the
present paper.

% %%%%%%%%%%%%%%%%%%%%%%%%%%%%%%%%%%%%%%%%%%%%%%%%%%%%%%%%%%%%%%%
\subsection{Impulse Approximation}

Basing on the discussion given above we consider here the nuclear
Compton amplitude in the impulse approximation (IA), Fig.\ref{fig1}.
Our aim here is to discuss the IA in terms of a covariant framework
valid at any momentum transfer and eventually develop a systematic
expansion scheme in terms of $Q^{-2}$ for the corrections due to
nuclear effects.

We take the imaginary part of
the nuclear Compton amplitude in IA and
find that the hadronic tensor of the nucleus with
momentum $P_A$ can be written as follows:
\begin{eqnarray}
\label{IA}
W_{\mu\nu}^A(P_A,q)
= \sum_{\tau=p,n} \int [dp]\, \text{Tr} \left[
  {\cal A}^\tau(p;P_A)\, \widehat{\cal W}^\tau_{\mu\nu}(p,q)
  \right],
\end{eqnarray}
where $\left[dp\right] \equiv d^4p / (2\pi)^4$ and $\widehat{\cal
W}^{p,n}_{\mu\nu}(p,q)$ is the hadronic tensor of the off-mass-shell
proton or neutron, given by Eqs.(\ref{Whatdef}) and (\ref{DirSt}).

The function ${\cal A}^\tau(p;P_A)$ is the proton
$(\tau=p)$ or the neutron $(\tau=n)$ propagator
in the nucleus which is defined as follows:%
\footnote{
To be more precise ${\cal A}(p;P_A)$
is the imaginary part of the ``hole" part
of the full nucleon propagator.}
\begin{eqnarray}
\label{A}
{\cal A}^\tau_{\alpha\beta}(p;P_A)
&=& \int d^3{\bf r}dt\,e^{ip_0t-i{\bf p\cdot r}}
\left\langle
P_A\left| \bar N_\beta^\tau(0) N_\alpha^\tau({\bf r},t) \right|P_A
\right\rangle.
\end{eqnarray}
Here we describe nucleons by the field operator
$N_\alpha^\tau({\bf r},t)$,
the averaging is taken over the nuclear state with momentum
$P_A$.

The expression for the nuclear tensor in Eq.(\ref{IA})
is covariant and can be evaluated in any reference frame.
It will be convenient, however, to work in the target rest frame,
in which the target four-momentum is $P_A=(M_A;{\bf 0})$,
and the momentum transfer to the nucleus,
$q=(q_0;{\bf 0}_\perp,-|{\bf q}|)$, defines the $z$-axis.
In order to extract the structure functions from the equation
for the amplitude, Eq.(\ref{IA}), we consider the components
of the hadronic tensor.
One can easily find from Eq.(\ref{W3}) that the structure
function $F_3$ is determined by the $xy$-component
of the hadronic tensor in the laboratory frame:
\begin{eqnarray}
\label{F3}
{\gamma} F_3 &=& 2i\,W_{xy},
\end{eqnarray}
where by ${\gamma}$ we denote the ratio of the space to time componets
of the momentum transfer,
\begin{eqnarray}\label{flux}
{\gamma}={|{\bf q}|\over q_0}=
\left(1 + {4M^2x^2\over Q^2}\right)^{1/2},
\end{eqnarray}
with $x=Q^2/2Mq_0$. Using Eq.(\ref{F3})
we extract the nuclear $F_3$  from
Eq.(\ref{IA}) and find,
\begin{eqnarray}
x_A F_3^A(x_A,Q^2)
&=& \frac1{2M_A}\sum_{\tau=p,n}\int [dp]\, x'\, \text{Tr}
    \left[
    {\cal A}^{\tau}(p)\,
    \left(\widehat{\cal G}_0^\tau(p,q) +
                \widehat{\cal G}_z^\tau(p,q)/\gamma \right)
    \right],
\label{F3A}
\end{eqnarray}
where $x_A=Q^2/2P_A\cdot q$ and $x'=Q^2/2p \cdot q$ are the
Bjorken variables for the nucleus and the bound (off-mass-shell)
nucleon with four-momentum $p$, respectively.

We stress here that Eq.(\ref{F3A}) has been derived within
a fully relativistic approach and no approximations have been
made in terms of $Q^2$. A similar approach (though focused at
asymptotic $Q^2$) have been already applied
to analyze the nuclear charged-lepton
structure functions $F_{1,2}$\cite{KPW94}
as well as $g_{1,2}$\cite{KMPW95}.
One important observation which follows
from Eq.(\ref{F3A}) is that, in general,
traces do not factorize into
completely separate nuclear and nucleon parts: the functions
$f^{(i)}$, which correspond to different Dirac structures
in Eq.(\ref{DirSt}),
enter $F_3^A$ with correspondingly different coefficients.
We note also that these functions depend in general on three variables:
$x',\ Q^2$ and $p^2$.
One arrives therefore to the conclusion that even in the impulse
approximation nuclear structure functions are not determined by those
of the on-mass-shell nucleon but are sensitive to their off-shell
behavior.

However in practice it may be quite sufficient to treat nuclei
as nonrelativistic systems.
The analysis of Refs.\cite{KPW94,KMPW95} shows
that both for spin independent ($F_1$ and $F_2$)
as well as for the spin ($g_1$ and $g_2$)
nuclear structure functions one can in fact recover
factorization in the nonrelativistic limit.%
\footnote{
We note however the effect of mixing of the spin structure functions
$g_1$ and $g_2$ observed in Ref.\cite{KMPW95}.
}
In this limit
nuclear structure functions factorize into convolution of
nonrelativistic nuclear spectral function and
``off-shell nucleon structure functions".
The latter are the definite combinations of
off-shell inelastic formfactors
(the functions $f^{(i)}$ in the discussed case)
with a correct on-shell limit.

In the next section we perform a similar nonrelativistic reduction
of relativistic expressions in Eq.(\ref{F3A}) to establish whether
a similar factorization holds for the structure function $F_3$.
An important point of the present analysis is that it can be
applied at arbitrary $Q^2$.

% %%%%%%%%%%%%%%%%%%%%%%%%%%%%%%%%%%%%%%%%%%%%%%%%%%%%%%%%%%%%%%%
\subsection{Nuclear Structure Functions in the Non-Relativistic Limit}
\label{nr-limit}

Our basic assumption in the remainder of the paper is that the
nucleus is a non-relativistic system, made up of weakly bound
nucleons.
This necessarily involves neglecting antinucleon degrees of freedom,
and corresponds to bound nucleons in the nucleus being slow,
$|{\bf p}|\ll M,\ |p_0-M|\ll M$.
In order to find nonrelativistic approximation to Eq.(\ref{F3A})
we have to
perform a nonrelativistic reduction of all Dirac structures in
the nucleon hadronic tensor.
In doing so we
follow the procedure which is described in detail in
Ref.\cite{Ku89,KPW94,KMPW95}.
Here we outline the basic steps of the method.
The essential relation is the relation between the relativistic
four-component nucleon field $N$ and the nonrelativistic two-component
operator $\psi$ (for the brevity of notations we suppress the isospin
index $\tau$),
\begin{equation}
\label{N-psi}
N({\bf p},t) = e^{-iMt}
\left(
	\begin{array}{r}
		Z\,\psi({\bf p},t) \\
\frac{
{\mbox{\boldmath$\sigma\cdot$}\displaystyle\bf p}
  }{\displaystyle 2M}\,\psi({\bf p},t)
	\end{array}
\right),
\end{equation}
where the nucleon operators are taken in a mixed $({\bf p},t)$
representation and
the renormalization operator $Z=1-{\bf p}^2/8M^2$
guarantees baryon number conservation.
We note here that Eq.(\ref{N-psi}) is valid to order $1/M^2$
in the series of $1/M$-expansion
and to this order does not depend explicitly on nuclear interactions
(see discussion in Appendix of Ref.\cite{KMPW95}).
We write the four-momentum of the bound nucleon as
$p=(M+\varepsilon, {\bf p})$ and introduce
the non-relativistic spectral function as follows:
\begin{equation}
\label{ANR}
{\cal P}(\varepsilon,{\bf p})
= \int dt\,
    e^{i\varepsilon t}
    \left\langle P_A|
        \psi^{\dagger}({\bf p},0)\psi({\bf p},t)
    |P_A\right\rangle/\left\langle P_A|P_A \right\rangle .
\end{equation}
Notice that, as it follows from its definition, the spectral function
is normalized to the number of protons ($Z$) and netrons ($N$) in the
nucleus:
\begin{eqnarray}
\label{norma}
\int{d\varepsilon d^3p\over (2\pi)^4}\,
{\cal P}^{p,n}(\varepsilon,{\bf p}) = (Z,N).
\end{eqnarray}

By substituting Eq.(\ref{N-psi}) into Eq.(\ref{A}) we find the relation
between relativistic and nonrelativistic spectral functions and then
reduce the four-dimensional Dirac basis to the two-dimensional spin
matrices.
We skip here technical details and summarize the results.
The non-relativistic limit of
traces with the components of the
operator $\widehat{\cal G}^\beta(p,q)$
can be written as follows:
\begin{mathletters}%
\label{NRtr}
\begin{eqnarray}
\frac1{2M_A}{\rm Tr}
\left[{\cal A}(p)\:\widehat{\cal G}_0(p,q)\right] &=&
        {\cal P}(\varepsilon,{\bf p})\:
        F_3(x',Q^2;p^2),\\
\frac1{2M_A}{\rm Tr}
\left[{\cal A}(p)\:\widehat{\cal G}_n(p,q)\right] &=&
        {p_n\over M}
        {\cal P}(\varepsilon,{\bf p})\:
        F_3(x',Q^2;p^2),
\end{eqnarray}
\end{mathletters}%
where $n=x,y,z$ stands for the spatial components and
$F_3(x',Q^2;p^2)$ is defined as follows:
\begin{eqnarray}\label{F3off}
 F_3(x',Q^2;p^2) &=&
f^{(0)}\left(1+\frac{p^2-M^2}{2M^2}\right) +
f^{(p)}\left(1+\frac{p^2-M^2}{M^2}\right) +
f^{(q)} + f^{(\gamma)} ,
\end{eqnarray}
where for the brevety of notations we omit explicit dependence of
functions $f_i$ on $x',\ Q^2$ and $p^2$.
Note also that $F_3$ in Eqs.(\ref{NRtr}) depends only on invariant
variables.

Few comments on Eqs.(\ref{NRtr}) are in order.  We emphasize here
that non-relativistic limit is taken with respect to the nucleon
momentum.  In derivation of Eqs.(\ref{NRtr}) we keep terms to order
${\bf p}^2/M^2$ and $\varepsilon/M$ and neglect terms of higher
orders.  At the same time Eqs.(\ref{NRtr}) are valid for arbitrary
momentum transfer $q$.

Eqs.(\ref{NRtr}) exhibit factorization of high-energy (structure
function $F_3$) and low-energy (spectral function ${\cal P}$)
domains.  We observe from (\ref{NRtr}) that instead of {\em four}
off-shell structure functions $f^{(i)}$, in non-relativistic domain
we deal only with {\em one} specific combination of them,
Eq.(\ref{F3off}), which enters the nuclear structure function.  We
assume that Eq.(\ref{F3off}) defines an analytical continuation of
the nucleon structure function $F_3$ to the off-shell region in the
vicinity of the mass-shell.  It is easy to see that at $p^2=M^2$
Eq.(\ref{F3off}) reduces to Eq.(\ref{F3on}) that assures a correct
on-shell limit of our analytic continuation of the structure
function.

%%..................................................................
\subsection{Generalized Convolution At Asymptotic and Finite $Q^2$}
\label{sf_nr}

The factorization property of Eqs.(\ref{NRtr}) allows us to write
the nuclear structure function as a convolution of the off-shell nucleon
structure function (\ref{F3off}) and the nuclear spectral function
(\ref{ANR}). We substitute Eq.(\ref{NRtr}) to Eq.(\ref{IA})
and obtain the desired connection:
\begin{eqnarray}
\label{IAnr}
x\,F_3^A(x,Q^2) =
\sum_{\tau=p,n}\int{d\varepsilon d^3p\over (2\pi)^4}
{\cal P}^\tau(\varepsilon,{\bf p})
        \left(1+\frac{p_z}{{\gamma} M}\right)
x'\,F_3^\tau(x',Q^2;p^2) ,
\end{eqnarray}
where we consider the nuclear $xF_3$ as a function of standard
Bjorken variable $x=Q^2/2Mq_0=x_A M_A/M$ and $\gamma$ is given
by Eq.(\ref{flux}). The variables which the off-shell nucleon
structure function depends on read
%%\vspace*{-\baselineskip}\vspace*{-\abovedisplayskip}
%
\begin{eqnarray}
\label{xprime}
x' &=& {x \over 1+(\varepsilon+{\gamma} p_z)/M},\\
p^2&=&M^2+2M\varepsilon -{\bf p}^2 .
\end{eqnarray}

We discuss now Eq.(\ref{IAnr}) in more detail. We recall that no
approximation is done with respect to $Q^2$, so that Eq.(\ref{IAnr})
is valid, in general, for any $Q^2$. Let us consider first the limit
of large
$Q^2\gg M^2$. In this case we put $\gamma=1$ in Eq.(\ref{IAnr}) and
find that $xF_3$ is smeared in nuclei exactly like $F_2$\cite{KPW94}.
However this similarity does not hold anymore when we study effects
due to finite $Q^2$.
%%\footnote{
%%Work on nuclear power corrections for the structure functions
%%$F_{1,2}$ as well as for polarized structure functions $g_{1,2}$ is
%%in progress.
%%}

To explore $Q^2$ dependent nuclear effects in Eq.(\ref{IAnr})
a useful observation is that all these effects come through
the parameter $\gamma$.
For the sake of convenience we consider the r.h.s. of
Eq.(\ref{IAnr}) as a function of $\gamma^2$. Then expanding the
under-integral expression in Eq.(\ref{IAnr}) about $\gamma^2=1$ in
Taylor series in $\gamma^2-1$ we obtain exactly the series in
$Q^{-2}$.  We skip here algebra details and show the resulting
expansion up to $Q^{-4}$ term,
\begin{eqnarray}
\label{q2-exp}
xF_3^A(x,Q^2)/A &\approx&
\left\langle\left(1+{p_z\over M}\right)
        x'F_3(x',Q^2;p^2)\right\rangle \nonumber\\
&&- {2M^2x^2\over Q^2}
        \left\langle {p_z\over M}
        \partial_{x'}\left(x'^2F_3(x',Q^2;p^2)\right)
        \right\rangle    \nonumber\\
&&+2\left({M^2x^2\over Q^2}\right)^2
        \left\langle
        {p_z\over M}G(x',Q^2;p^2)
        +{p_z^2\over M^2}x'\partial_{x'}G(x',Q^2;p^2)
        \right\rangle .
\end{eqnarray}
Here to simplify the notations the brackets $\langle\cdots\rangle$ denote
the averaging over the nuclear spectral function,
\begin{eqnarray}
\label{ave}
    \langle{\cal O}\rangle &=&
        {1\over A}\int{d\varepsilon d^3p\over (2\pi)^4}\,
        {\cal P}(\varepsilon,{\bf p})\:{\cal O}(\varepsilon,{\bf p}) .
\end{eqnarray}
We also do not write explicitly the sum over the nucleon isospin states
in Eqs.(\ref{q2-exp},\ref{ave}).
Note that $x'$ in Eq.(\ref{q2-exp}) is given by Eq.(\ref{xprime})
with $\gamma=1$, and $\partial_{x}$ denotes the derivative with respect
to $x$. The function $G(x,Q^2;p^2)$ stays for
the following combination of $F_3$ and its derivative,
\begin{eqnarray}
\label{G}
G(x,Q^2;p^2)=4xF_3(x,Q^2;p^2)+x^2\partial_x F_3(x,Q^2;p^2).
\end{eqnarray}
Few comments on Eq.(\ref{q2-exp}) are in order. A natural question to ask
is:  What is an effective expansion parameter there? Formally this is
$M^2x^2/Q^2$. However this parameter is always multiplied by
nuclear averages which involve powers of the nucleon
momentum $p_z$ which is small in the nucleon mass scale. It may
happen that this effect compensates a large factor $M^2$. In fact it
does.  One can see this if one treats the Fermi motion and binding
effects in Eq.(\ref{q2-exp}) perturbatively, i.e. by expanding the
structure function in $(\varepsilon+p_z)/M$ about $x'=x$ (detailed
discussion of the expansion procedure see in Sec.\ref{D-expan}).
Then we find that the nuclear matrix element of $Q^{-2}$ term is of
order $\langle{\bf p}^2\rangle/M^2$.  For the $Q^{-4}$ term in the
expansion (\ref{q2-exp}) we find that two terms in the matrix element
cancel each other to order ${\bf p}^2/M^2$ yielding the finite
contribution only in order ${\bf p}^4/M^4$.  We conclude from
this discussion that an effective expansion parameter in
Eq.(\ref{q2-exp}) is ${\bf p}^2x^2/Q^2$.

We recall in this respect that Eq.(\ref{IAnr}) itself valid to order
$\varepsilon/M\sim{\bf p}^2/M^2$.  One must comment therefore that
the nuclear $Q^{-4}$ correction can not be calculated consistently
without considering relativistic corrections to order ${\bf p}^4/M^4$
to Eq.(\ref{IAnr}).

%%%%%%%%%%%%%%%%%%%%%%%%%%%%%%%%%%%%%%%%%%%%%%%%%%%%%%%%%%%%%%%%%%%%%%%%%
\section{Off-shell Nucleon Structure Functions}
\label{offshell}

In order to obtain an estimate of off-shell effects in the nucleon
structure functions we consider the latter in the leading order
(impulse approximation for quarks, see Fig.\ref{fig2}).
In analysing this
diagram we consider the target momentum $p$ to be generally
off-mass-shell. As a convenient tool we use the Sudakov variables,
i.e.  parameterize the quark momentum in the loop, $k$, in terms of
external momenta $p$ and $q$,
\begin{eqnarray}
\label{sudakov}
k=\alpha p+\beta q'+k_\perp,
\end{eqnarray}
where $q'=q+\xi p$ and $k_\perp$ is two-dimensional vector (with
$k_\perp^2<0$) perpendicular to both $p$ and $q$. We choose the
quantity $\xi$ so that $q'^2=0$ that gives
\begin{eqnarray}
\label{xi}
\xi={2x\over 1+(1+4x^2p^2/Q^2)^{1/2}}.
\end{eqnarray}
Note that $\xi$ is in fact the Nachtmann variable for the target
with the mass squared $p^2$.

In doing the integration with respect to $\beta$ we assume that the
quark correlator has standard analytic properties in terms of the
variables $s=(k-p)^2$ and $k^2$, i.e. it is an analytic function in
complex $s$- and $k^2$-planes with the cuts for $s>0$ and $k^2>0$.
Using the dispersion relation for the quark correlator we can write
the structure function as follows,
\begin{eqnarray}
\label{F3:spectral}
F_3(x,Q^2;p^2) = \int {d\alpha ds d^2k_\perp \over 1-\alpha}
\rho(\alpha,s,k^2;p^2)\, 2p\!\cdot\!q\,\delta\left((k+q)^2\right),
\end{eqnarray}
where $0\le\alpha\le 1$ and
the integration over $s$ runs over the spectrum of the
``spectator" quark system and $\rho$ describes the spectrum of
these states.%
\footnote{
To simplify notations we drop the explicit dependence of $\rho$
on a the normalization point $\mu^2$ which is fixed here to be $Q^2$.
}
The squared quark momentum $k^2$ in Eq.(\ref{F3:spectral}) is
\begin{eqnarray}
\label{ksq}
k^2=\alpha\left({s\over \alpha-1}+p^2\right)
         + {k_\perp^2\over 1-\alpha} .
\end{eqnarray}

At asymptotic $Q^2\gg 1\,$GeV$^2$ one can neglect the quark virtuality
$k^2$ (which is assumed to be finite) in the $\delta$-function in
Eq.(\ref{F3:spectral}) and the latter then goes to
$\delta(\alpha-x)$, so that we find in this limit:
\begin{eqnarray}
\label{F3:as}
F_3^{\text{as}}(x;p^2) = \int {ds d^2k_\perp \over 1-x}
\rho(x,s,k^2;p^2).
\end{eqnarray}
Now, we observe from Eq.(\ref{F3:as}) that $p^2$ dependence of the
structure functions has two primary sources: explicit dependence of
the quark spectral density on $p^2$ (``dynamical" $p^2$ dependence)
as well as $p^2$ dependence of the quark off-shellness $k^2$
(``kinematical" $p^2$ dependence).
Our general idea here is to fix the overall $p^2$ dependence of the
structure functions by requiring the conservation of the
normalization of $F_3$ off-shell.

We recall in this respect that in the parton model $F_3$ is
interpreted as the valence quarks distribution in the target. The
integral of $F_3$,
\begin{eqnarray}
\label{GLS_def}
S_{\text{GLS}}(Q^2) &=& \int_0^1 dx\, F_3(x,Q^2),
\end{eqnarray}
measures at asymptotic $Q^2$ the number of valence quarks (or the
baryon charge) in the target, i.e. for the nucleon $S_{\text{GLS}}=3$
(the Gross-Llewellyn-Smith (GLS) sum rule\cite{GLS69}).  The baryon
current is conserved by strong interaction and, as a working
hypothesis, we assume that $S_{\text{GLS}}$ is not renormalized
off-shell (at least in the vicinity of $p^2=M^2$). In the vicinity of
$p^2=M^2$ one can write this condition as follows:
\begin{eqnarray}
\label{nonrenorm}
\int_0^1 dx\,\partial_{p^2} F_3^{\text{as}}(x;p^2) = 0,
\end{eqnarray}
where $\partial_{p^2}$ denotes the derivative with respect to $p^2$.
In order to make use of this condition we will follow
Ref.\cite{KPW94} and assume that the spectrum in $s$ can be
approximated by that calculated for a single effective mass $\bar s$
and write the spectral density as
\begin{eqnarray} \label{rho}
\rho = \Phi(k^2,p^2) \delta(s-\bar s) .
\end{eqnarray}
This model spectral density has been used in Ref.\cite{KPW94} to
describe the valence quark distribution in the proton. It was found
that the $k^2$ dependence of (\ref{rho}) taken in the form
$(1-k^2/\Lambda^2)^{-n}$ with the exponent $n=4$ and the cutoff
$\Lambda^2=1.2\ \text{GeV}^2$ leads to a good description
of the data at high $Q^2$. To be specific we note here that
these parameters have been fixed from the fit to
the nucleon structure function data at average momentum transfer
$Q^2\approx 5-10\ \text{GeV}^2$, which fixes the normalization point
for the spectral density. The leading $Q^2$ dependence
of the structure functions can then be obtained
by applying the evolution equation.

Extrapolating $\rho$ off-mass-shell we consider two possible models
for $p^2$-dependence in Eq.(\ref{rho}).  In the first model
\cite{KMPW95} we assume that $p^2$ and $k^2$ dependencies factorize,
$\Phi(k^2,p^2)=\varphi(p^2)\Phi(k^2)$.  Then the function
$\varphi(p^2)$ can be determined from Eq.(\ref{nonrenorm}).  In the
second model \cite{KPW94} we assume that the $p^2$ dependence comes
from that of the cutoff, $\Lambda=\Lambda(p^2)$. Its $p^2$ dependence
is determined again from Eq.(\ref{nonrenorm}).

Since we are interested in off-shell behavior of the structure
functions in the vicinity of the mass-shell, the function of interest
is $\partial_{p^2} F_3$ taken at $p^2=M^2$. We found that both models
give similar results for this function.  Fig.\ref{fig_f3} and
Table \ref{tablo} give an idea of the magnitude of the off-shell
effect in the structure function $F_3$.

Going down to finite $Q^2$ we should comment that there is no direct
reason to expect that Eq.(\ref{nonrenorm}) will remain valid, since
$S_{\text{GLS}}$ does not correspond to the matrix element
of a conserved current at finite $Q^2$.
Finite $Q^2$ corrections can be of two different kinds:
dynamical ones, which are due to final state interaction of the struck
quark with the spectator quark system; and kinematical corrections
which are due to finite mass of the target.
We focus here on the discussion of $p^2$ dependence of the structure
functions. It is likely that the dynamical power corrections depend
weakly on the target mass and we do not consider these effects here.
In order to evaluate corrections due to finite mass of the target,
let us consider the argument of the $\delta$-function
in Eq.(\ref{F3:spectral}) in more detail.
Using Eq.(\ref{sudakov}) one can write it as follows,
\begin{eqnarray}
\label{arg}
(k+q)^2 = 2p\!\cdot\!q'\,(\alpha-\xi) + (k-\xi p)^2.
\end{eqnarray}
The last term in this equation measures the contribution
from non-collinear components in Eq.(\ref{sudakov}).
Next we expand the $\delta$-function in Eq.(\ref{F3:spectral})
in power series in $(k-\xi p)^2$. The zero order term in this expansion
picks $\alpha=\xi$
and we find the following contribution to $F_3$,
\begin{eqnarray}
\label{xi-scaling}
F_3^{\text{as}}(\xi;p^2)\,
{1-\xi^2 p^2/Q^2\over 1+\xi^2 p^2/Q^2}.
\end{eqnarray}
We found that the first order correction in $(k-\xi p)^2$ vanishes if
the spectral density $\rho$ does not depend on $\alpha$ explicitly
(implicitly we have already used this assumption in Eq.(\ref{rho})).
This can be seen after changing the integration $d^2k_\perp\to dk^2$
in Eq.(\ref{F3:spectral}) and writing $(k-\xi p)^2$ in terms of the
variables $k^2$ and $s$.  We find therefore that the corrections due to
non-collinear components in the $\delta$-function in
Eq.(\ref{F3:spectral}) appear starting from the order $Q^{-4}$ and
higher.  It is then justified to use Eq.(\ref{xi-scaling}) to order
$Q^{-2}$ (though higher order terms could be important at $x\to1$).
To this end we expand Eq.(\ref{xi-scaling}) in
series in $p^2/Q^2$ and keep only the first order correction. We then
have,
\begin{eqnarray}
\label{F3:tmc}
F_3(x,Q^2;p^2)=F_3^{\text{as}}(x;p^2)
- {p^2\over Q^2}x\partial_x
\left(x^2 F_3^{\text{as}}(x;p^2)\right)
\end{eqnarray}

We find from the latter equation that to order $Q^{-2}$
the right-hand side of
Eq.(\ref{F3:tmc}) is expressed entirely in terms of the asymptotic
part of $F_3$. This observation allows us to calculate the derivative
$\partial_{p^2} F_3$ to order $Q^{-2}$ using the asymptotic
structure function given by Eqs.(\ref{F3:as},\ref{rho}).

%%%%%%%%%%%%%%%%%%%%%%%%%%%%%%%%%%%%%%%%%%%%%%%%%%%%%%%%%%%%%%%%%%%%%%%%%
\section{Nuclear Effects in Sum Rules}
\label{sumrule}

The moments of the structure functions are determined by the matrix
elements of quark-gluon operators in the target. Those moments which
correspond to the matrix elements of conserved currents are of
special interest. We recall that $F_3$ obeys to the
Gross-Llewellyn-Smith sum rule\cite{GLS69}, i.e. the GLS integral
Eq.(\ref{GLS_def}) measures the baryon charge of the target. We
stress however that this correspondence is approximate and exists
only at asymptotic $Q^2$. At finite $Q^2$ the GLS sum rule receives
the QCD corrections.
The following equation
summarizes the theoretical
status of perturbative as well as power twist-4 corrections to the
GLS sum rule for the proton,
\begin{eqnarray}\label{GLS_QCD}
S_{\text{GLS}}^p(Q^2) &=& 3\left[
        1-{\alpha_s\over\pi}
        -3.25\, \left({\alpha_s\over\pi}\right)^2
        -31.06\, \left({\alpha_s\over\pi}\right)^3 +\cdots \right.\nonumber\\
&&\left.  -\frac8{27}{\langle\langle{\cal O}\rangle\rangle\over Q^2}
\right]+{M^2\over Q^2}\int_0^1dx\,x^2 F_3^p(x,Q^2),
\end{eqnarray}
where $\alpha_s=\alpha_s(Q^2)$ is the running QCD coupling
constant. The coefficients of $\alpha_s^2$
and $\alpha_s^3$ have been calculated in \cite{GLS_cor}
(the coefficients in Eq.(\ref{GLS_QCD}) are given for $N_f=4$).
A power correction in Eq.(\ref{GLS_QCD}) due to twist-4 operators has
been evaluated in \cite{BrKo87,RR94} by the QCD sum rules method.
Ref.\cite{BrKo87} gives $\langle\langle{\cal O}\rangle\rangle \approx
0.33\,$GeV$^2$ (with a typical QCD sum rules error about 30\%).
The estimate of Ref.\cite{RR94} is slightly larger,
$\langle\langle{\cal O}\rangle\rangle\approx0.53\,$GeV$^2$.
As the last term in Eq.(\ref{GLS_QCD})
we write also explicitly the target mass correction.
One observes from Eq.(\ref{GLS_QCD}) a partial cancellation
of the twist-4 power correction and the target mass term. We take
$\langle\langle{\cal O}\rangle\rangle$ of Ref.\cite{RR94}
to obtain an upper estimate for the $Q^{-2}$ term in Eq.(\ref{GLS_QCD}),
then we have
$\delta S_{\text{GLS}}/3=-0.11\,\text{GeV}^2/Q^2$.

Now we calculate nuclear correction to the GLS sum rule by integrating
Eq.(\ref{q2-exp}) over $x$.  In the right-hand side of
Eq.(\ref{q2-exp}) we change the integration variable $x\to x'$ and
after some algebra we have:
\begin{eqnarray}
\label{GLS_A}
S^A_{\text{GLS}}(Q^2) &=&
\left\langle
\int_0^1 dx'\left(
        1
        +\frac{4\, x'^2 p_z^2}{Q^2}
        +{8\,x'^4 p_z^2(M\varepsilon+5p_z^2)\over Q^4}
\right)
F_3^N(x',Q^2;p^2)
\right\rangle .
\end{eqnarray}
We observe from Eq.(\ref{GLS_A}) that nuclear corrections cancel out
%%to order ${\bf p}^2/M^2$
in the leading term in Eq.(\ref{q2-exp}) and we find this not surprising.
Thus corrections appear starting from $Q^{-2}$ order.
We write explicitly also the $Q^{-4}$ term just to illustrate our general
conclusion about the expansion parameter in nuclear $Q^{-2}$ series
(see discussion at the end of Sec.\ref{sf_nr}).
This term introduces an extremely small correction and we do not
consider it in what follows.

Note that the overall nuclear $Q^{-2}$ correction in Eq.(\ref{GLS_A}) comes
from two sources:
 1) the second term in the r.h.s. of
Eq.(\ref{GLS_A}) which is due to Fermi-motion of bound nucleons,
and 2) $p^2$ dependent $Q^{-2}$ terms in the
off-shell nucleon structure function, Eq.(\ref{F3:tmc}).
In order to evaluate the latter we observe that,
since an average deviation from the mass shell in Eq.(\ref{GLS_A}) is
small, we can expand the off-shell nucleon structure function in
$\Delta p^2=p^2-M^2$ about the mass shell point and keep
only the first order correction.
Using Eq.(\ref{GLS_A}) and Eq.(\ref{F3:tmc}) we find,
\begin{eqnarray}
\frac1A S_{\text{GLS}}^A(Q^2) &=& S_{\text{GLS}}^N(Q^2)
+{\langle \Delta p^2+\frac43{\bf p}^2\rangle \over Q^2}
        \int_0^1 dx\,x^2 F_3^N(x,Q^2) \nonumber\\
&&
+{\langle \Delta p^2\rangle \over Q^2}
        \int_0^1 dx\,x^2 p^2\partial_{p^2} F_3^N(x,Q^2),
\label{GLS_A2}
\end{eqnarray}
where we have assumed $N=Z$ in the nucleus and
$F_3^N=\frac12(F_3^p+F_3^n)$ and
$S_{\text{GLS}}^N$ are respectively the structure function
and the GLS integral for the on-mass-shell isoscalar
nucleon. The derivative with respect to $p^2$
in the last term of Eq.(\ref{GLS_A2}) is
taken at the mass shell point, $p^2=M^2$.
The average off-mass-shell shift for the bound nucleon,
$\langle\Delta p^2\rangle=\langle 2M\varepsilon-{\bf p}^2\rangle$,
mean nucleon separation energy, $\langle\varepsilon\rangle$, and mean
nucleon momentum squared, $\langle{\bf p}^2\rangle$, are determined by
averaging with the nuclear spectral function by Eq.(\ref{ave}).

In order to evaluate the size of nuclear corrections in
Eq.(\ref{GLS_A2}), we refer here the results of computations of the
spectral function for finite nuclei in the local density
approximation \cite{SFFB94,NDM94}.  In that approach the spectral
function is decomposed in terms of a single particle contribution
(mean field) and a correlation part. It is assumed that the latter is
insensitive to the finite size of nuclei and is given by nuclear
matter spectral function corrected with the help of local density
approximation. As a result one finds that for a
range of nuclei with $16\le A\le 208$ both the mean field and the
correlation part of kinetic energy amount approximately $17\,$MeV and
slowly increase with $A$\cite{NDM94}. The mean field and the
correlation parts of mean removal energies are about $30\,$MeV and
$20\,$MeV respectively, and both increase with $A$.  One of the
traditional targets for neutrino DIS is iron ${}^{56}$Fe.
For this nucleus, basing on the discussion given
in Ref.\cite{NDM94}, we take $\langle{\bf
p}^2\rangle/2M=35\,$MeV and $\langle\varepsilon\rangle=-56\,$MeV.
From these estimates we find
$\langle\Delta p^2\rangle_{\text{Fe}}=-0.17\,$GeV$^2$.
For comparision we give here also the corresponding values for the
deuteron nucleus,
$\langle\varepsilon\rangle_{\text{D}}=
\epsilon_{\text{D}}-\langle{\bf p}^2\rangle_{\text{D}}/2M$,
where $\epsilon_{\text{D}}=-2.2\,$MeV is the deuteron binding energy. Taking
the Bonn potential deuteron wave function we have
$\langle{\bf p}^2\rangle_{\text{D}}/2M\approx 7\,$MeV and
$\langle\Delta p^2\rangle_{\text{D}}=-0.03\,$GeV$^2$.
Using these estimates as well as the results of Sec.\ref{offshell} we
find from Eq.(\ref{GLS_A2}) for the iron and deuteron nuclei,
\begin{mathletters}%
\label{dS}
\begin{eqnarray}
{\delta S_{\text{GLS}}^{\text{Fe}}\over 3} &=&
- {4.0\cdot10^{-3}\over Q^2},\\
{\delta S_{\text{GLS}}^{\text{D}}\over 3}  &=&
- {6.3\cdot10^{-4}\over Q^2},
\end{eqnarray}
\end{mathletters}%
where the coefficients are taken in units of GeV$^2$
($\delta S_{\text{GLS}}$ in Eq.(\ref{dS}) denotes the
nuclear correction to $S_{\text{GLS}}^N$
in Eq.(\ref{GLS_A2}), the factor $1/3$ is introduced
for the sake of convenience).

It is useful to compare the sizes of different contributions
to Eq.(\ref{dS}). The main contributions come from the second
term in the r.h.s. of Eq.(\ref{GLS_A2}) which consists from the Fermi-motion
term and the target mass correction term
which are of opposite sign.
Since the average off-mass-shell shift, $\langle\Delta p^2\rangle$,
is negative and about as twice as large as the Fermi-motion term,
$\langle\frac43{\bf p}^2\rangle$, the overall contribution becomes negative.
The last term in Eq.(\ref{GLS_A2}), which describes the deformation of
the nucleon parton distributions due to off-mass-shell shift,
brings about 20\% of the total contribution to Eq.(\ref{dS}).
Note that if we disregard the effect of $p^2$-dependence of the
bound nucleon structure function (which is a standard assumption
in many convolution model calculations)
we would leave only with the Fermi-motion
correction which is about of the same size as qouted in Eq.(\ref{dS})
but of {\em opposite} sign.

We conclude from the present analysis that nuclear corrections to the
GLS sum rule cancel out in the leading order in $Q^{-2}$ series.
A direct reason for
this is that at asymptotic $Q^2$ the GLS sum rule corresponds to the
baryon number of the target which is  not renormalized by strong
interaction. The GLS sum rule receives a tiny nuclear
correction in $Q^{-2}$ order which turns out to be
more than an order of magnitude less than the corresponding QCD
power correction.

To conclude this section we give an expression for the arbitrary
moment of nuclear structure function which follows from
Eq.(\ref{q2-exp}). We integrate both sides of this equation with the
weight $x^{n-1}$ and keep terms to order ${\bf p}^2/M^2$. We have,
\begin{eqnarray}
M_n(Q^2)     &=& \int_0^1 dx\,x^n F_3(x,Q^2),\\
M_n^A(Q^2)/A &=& \left(
        1
        +{\langle\varepsilon\rangle\over M}\,n
        +{\langle{\bf p}^2\rangle\over 6M^2}\,n(n+1)
        \right)M_n^N(Q^2)   \nonumber\\
&& +\langle\Delta p^2\rangle\partial_{p^2}M_n^N(Q^2)
+{2\langle{\bf p}^2\rangle\over 3Q^2}\,(n+1)(n+2)M_{n+2}^N(Q^2) ,
\label{moments}
\end{eqnarray}
where $\partial_{p^2}M_n$ denotes the moments of $\partial_{p^2}F_3$
which can be calculated by Eq.(\ref{F3:tmc}) (see also Table \ref{tablo}).
We should comment however that application of Eq.(\ref{moments})
to large $n$ should be taken with care.
Indeed, Eq.(\ref{moments}) has been obtained by expanding
the expressions like $(1+(\varepsilon+p_z)/M)^n$ to order
$\varepsilon/M\sim p_z^2/M^2$.  Obviously we are not allowed to do
this when already the first order correction is of order of 1.  Using
the estimates for typical $\varepsilon$ and ${\bf p}^2$ from above we
see that this happens for $n\sim 10$ for heavy nuclei.
To calculate the moments with larger $n$ one has to use the full
nuclear spectral function.

% .....................................................................
\section{Derivative Expansion Formula}
\label{D-expan}

In this section we construct an approximation to Eq.(\ref{IAnr})
taking further the advantage of a small parameter ${\bf p}/M$.
The smearing
of the nucleon structure function in a nucleus is governed by the nuclear
spectral function ${\cal P}(\varepsilon,{\bf p})$ which describes the
distribution of bound nucleons over momenta and energies.  Since
characteristic momenta ${\bf p}$
and energies $\varepsilon$ are small as compared to the nucleon mass, an
average deviation of the bound nucleon Bjorken variable, $x'$, from the
nuclear one, $x$, is also small.
This simple observation allows us to obtain an approximate expression
for the nuclear structure function in Eq.(\ref{IAnr}) by expanding
$x'F_3(x',Q^2;p^2)$ in a Taylor series in ${\bf p}/M$ and $\varepsilon/M$
around $p=(M,{\bf 0})$.
In doing so we integrate the resulting expansion term by term and
keep terms up to order $\varepsilon/M$ and ${\bf p}^2/M^2$.
As a result we
obtain an approximation to Eq.(\ref{IAnr}) in terms of a simple
derivative expansion formula:
\begin{eqnarray}
\label{Dex}
xF^A_3(x,Q^2)/A &\simeq&
        xF_3^N(x,Q^2)
    -\frac{\langle\varepsilon\rangle}{M}\:
            x\partial_x{\left(xF_3^N(x,Q^2)\right)}
    +\frac{\langle{\bf p}^2\rangle}{6M^2}\:
            x^2\partial_x^2{\left(xF_3^N(x,Q^2)\right)} \nonumber\\
&&    +{\langle\Delta p^2\rangle}
            \partial_{p^2}\left(xF_3^N\right)(x,Q^2)
    +\frac23\,\frac{\langle{\bf  p}^2\rangle}{Q^2}\:
            x^3\partial_x^2{\left(x^2 F_3^N(x,Q^2)\right)} .
\end{eqnarray}
Here
we assume the isoscalar target with equal numbers of protons
and neutrons and
the coefficients are expressed in terms of averaged nucleon
separation energy and momentum squared, Eq.(\ref{ave}).

We note that Eq.(\ref{Dex}) being integrated over $x$ reproduces
Eq.(\ref{GLS_A2}) for the GLS sum rule as well as Eq.(\ref{moments})
for higher moments. In fact the inverse statement is also true,
Eq.(\ref{Dex}) can be derived from
Eq.(\ref{moments}) by applying the Mellin transformation.

The expansion formula (\ref{Dex}) is similar to those which has been used
in the discussion of the EMC ratio for the structure function $F_2$
\cite{Ku89,KPW94}, as well as in the analysis
of nuclear effects in the spin structure functions $g_1$ and
$g_2$\cite{KMPW95} at asymptotic $Q^2$. We note in this respect that
(\ref{Dex}) includes also $Q^{-2}$ corrections due to nuclear effects.

Figs.\ref{fig_ratio},\ref{fig_ratio_q}
show the ratio $R_3=F_3^A/AF_3^N$ calculated
by Eq.(\ref{Dex}) for the iron and deuterium nuclei for different
$Q^2$. Main features of the behavior of the ratio $R_3$ calculated
at large $Q^2$ without the off-shell term are similar to those of the
corresponding ratio for $F_2$ (EMC effect) and are determined by the
competition of the separation energy and the kinetic energy terms in
Eq.(\ref{Dex}). The position of the dip of $R_3$ and its slope at
large $x$ are governed by the values of $\langle\varepsilon\rangle$
and $\langle{\bf p}^2\rangle$.
The effect of the last term in Eq.(\ref{Dex}) is similar to that
of the kinetic energy term -- it gives a positive correction.
However this term is largely compensated by the off-shell term
in Eq.(\ref{Dex}).
We observe from Fig.\ref{fig_ratio} a sizable correction due to the
off-shell term. Its main effect is to increase the dip in $R_3(x)$
and move its position to a bit larger $x$. Fig.\ref{fig_ratio_q}
shows the $Q^2$ dependence of the ratio $R_3$. Note that the main
source of the $Q^2$ dependence of the ratio
is the off-shell term in Eq.(\ref{Dex}) which incorporates also
a target mass correction.
We observe therefore that the size of the off-shell effects in the nuclear
structure function even becomes larger the smaller $Q^2$.

We present also the estimates of nuclear corrections
to the GLS integral truncated from below
which is relevant in
view of the extraction of the sum rule value from the data.
Integrating Eq.(\ref{Dex}) we have:
\begin{eqnarray}
\label{GLS_x}
&& S_{\text{GLS}}(x,Q^2)     = \int_x^1 dy\, F_3(y,Q^2), \\
&& S_{\text{GLS}}^A(x,Q^2)/A = S_{\text{GLS}}^N(x,Q^2)
+\frac{\langle\varepsilon\rangle}{M}\:xF_3^N(x,Q^2)
        -\frac{\langle{\bf p}^2\rangle}{6M^2}\:
        x^2\partial_x F_3^N(x,Q^2) \nonumber\\
&& +{\langle\Delta p^2\rangle}
      \partial_{p^2} S_{\text{GLS}}^N(x,Q^2)
+\frac23\,\frac{\langle{\bf  p}^2\rangle}{Q^2}
\left( -x^4\partial_x F_3(x,Q^2) + 2\int_x^1 dy\,y^2 F_3(y,Q^2)\right),
\label{GLS_x_A}
\end{eqnarray}
where $\partial_{p^2}S_{\text{GLS}}$ denotes the GLS integral
of $\partial_{p^2}F_3$.
Fig.\ref{fig_GLS_x} shows the $x$ dependence of nuclear corrections
to the GLS integral.
We observe that while the corrections practically
cancel out at $x=0$, according to the discussion of Sec.\ref{sumrule},
these are finite and negative in the
region of practical interest and may reach up to $\sim 10\%$ for a heavy
nucleus if one truncates the integral at $x\sim 0.2$.

We conclude this section by discussing the region of applicability of
the expansion formula (\ref{Dex}) which is both $x$ and $Q^2$
dependent.  The expansion procedure which leads to Eq.(\ref{Dex}) can
be applied where the structure function is smooth.
This is true when we are outside the elastic and resonance region,
i.e. when $(p+q)^2\ge s$ with $s\sim 2\,$GeV$^2$.
The latter condition can be written in terms of the bound nucleon
Bjorken variable as $x'>x_{\text{th}}$, where
$x_{\text{th}}=(1+(s-M^2)/Q^2)^{-1}$ describes the production
threshold of the states with the invariant mass squared $s$.
The bound nucleon Bjorken variable $x'$ is smeared about the point
$x'=x$ by Eq.(\ref{xprime}). It is then clear that we do not get into
a dangerous region if,
\begin{eqnarray}
\label{bounds}
1 -x/x_{\text{th}} \gtrsim p_{\text{char}}/ M,
\end{eqnarray}
where $p_{\text{char}}$ is a characteristic momentum of the bound
nucleon.
At asymptotic $Q^2$, where $x_{\rm th}\to 1$, Eq.(\ref{bounds}) tells
us that the expansion formula Eq.(\ref{Dex}) should be a good
approximation for $x\lesssim 0.75$ for heavy nuclei (for the deuteron
the corresponding region is wider).
We see also that for finite $Q^2$ the region of applicability of
Eq.(\ref{Dex}) becomes tighter.
We note however that, as a matter of fact,
even for large $x$ outside the region
(\ref{bounds}) an approximate equation (\ref{Dex}) follows to the
exact calculation by Eq.(\ref{IAnr}) though slightly overestimates
the effect of Fermi-motion. This is not surprising because
thresholds effects in the structure function restrict the integration
region in Eq.(\ref{IAnr}).

%\newpage
%%%%%%%%%%%%%%%%%%%%%%%%%%%%%%%%%%%%%%%%%%%%%%%%%%%%%%%%%%%%%%%%%%%%%%%%%
\section{Summary and Conclusions}
\label{finita}

In the present paper we have analyzed nuclear effects in the structure
function $F_3$.
Our starting point was relativistic impulse approximation for the
nuclear Compton amplitude.

%% FACTORIZATION
An important feature of the impulse approximation is the
factorization property: nuclear amplitude is factored into the
nucleon Compton amplitude (generally off-shell) and a relativistic
nucleon spectral function.
However, because of off-shell effects, factorization between the
amplitudes is not translated automatically into factorization between
the structure functions, that is assumed in usual convolution models.
Indeed, as we discussed in the paper, the structure of the hadronic
tensor of the off-shell nucleon is more complicated than that of the
on-shell one.  A direct reason for this is that one can not use the
Dirac equation for the off-shell nucleon which greatly reduces the
number of independent amplitudes in the hadronic tensor.
In particular we found that the structure function $F_3$ splits into
four independent inelastic form-factors for the nucleon off-shell.
Similar observations have been already done for the structure
functions $F_2$\cite{KPW94} as well as for $g_1$ and
$g_2$\cite{KMPW95}.
Obviously these observations
make more complicated
the problem of extraction of nucleon structure functions from nuclear
data and even could introduce a principal uncertainty into this
problem.

%% 1/M EXPANSION AND NONRELATIVISTIC LIMIT

We found that the factorization for the structure functions is
remarkably recovered in the limit of weak nuclear binding
when a nucleus is considered to be a nonrelativistic system.
In order to observe this we have done
a systematic $1/M$-expansion of nuclear matrix elements which enter
the nuclear Compton amplitude and found that
to order $1/M^2$ the nuclear structure function can be expressed in
terms of the generalized convolution of the (non-relativistic)
nuclear spectral function and a definite combination of the
off-shell nucleon inelastic form factors which we take for the
definition of the ``off-shell nucleon structure function".
We note that this observation is generic for all
structure functions \cite{KPW94,KMPW95}.

%% 1/Q^2 SERIES
We stress here that our analysis is valid for arbitrary momentum
transfer squared $Q^2$.
As a practical output of our approach we have developed a systematic
expansion scheme of nuclear structure function in ``nuclear twist''
series, i.e. power $Q^{-2}$ corrections caused by nuclear effects.
We argued that an effective parameter in this expansion is
$p_{\text{char}}^2x^2/Q^2$ with $p_{\text{char}}$ being the
characteristic momentum of the bound nucleon.
This allows us to apply the $Q^{-2}$ expansion down to small $Q^2\sim
0.1\,$GeV$^2$ and justifies to keep only $Q^{-2}$ correction at
$Q^2\gtrsim 1\,$GeV$^2$.

One should comment however that these observations are done within
the impulse approximation and other sources of nuclear effects,
such as the FSI, might be important in $Q^{-2}$ order.
It remains a challenging problem to quantify the FSI
corrections in nuclear DIS.

%% SUM RULES
We applied our approach to calculate the moments
of the nuclear structure function and in particular
the Gross-Llewellyn-Smith
sum rule. Nuclear corrections to the GLS sum rule cancel out
in the leading order, which is due to the baryon charge
conservation.
We have calculated also nuclear $Q^{-2}$ correction to the GLS sum rule
which turned out to be negative and small (more than an order of
magnitude less than the corresponding correction to the nucleon
structure function itself as estimated by the QCD sum rules).
Notice however that nuclear corrections are sizable for truncated
GLS integrals that may have an impact on extraction of the value
of the nucleon GLS sum rule from the data.

%% OFF-SHELL
We note here that in our analysis we can not avoid
the problem of off-shell behavior of the amplitudes and have to deal with the
problem explicitly. Unfortunately a rigorous approach, which would allow us
to calculate the structure functions off-shell, is lacking and we
have to relay more on model assumptions.
In the present paper we evaluate off-shell effects basing on the
non-perturbative parton model and assuming that the
nucleon GLS integral is not renormalized off-shell, which is
motivated by the baryon charge conservation.
The latter condition allows us to evaluate the off-shell behavior of
$F_3$ at large $Q^2$ while the consideration of the target mass corrections
leads to an estimate of this effect at finite $Q^2$ as well.

We found a sizable impact of off-shell effects on higher moments of
the nuclear structure function and on its $x$ dependence.
Talking about the ratio of nuclear to nucleon structure functions $R_3$
we observe that the off-shell effects tend to increase the dip at $x\sim 0.6$.
This effect is even more prominent at finite $Q^2$.
In view of this observation
%%(and having in mind the EMC effect)
we notice the importance of the corresponding finite $Q^2$ analysis
for the structure function $F_2$ as well as for the spin
structure functions $g_1$ and $g_2$.

%%%%%%%%%%%%%%%%%%%%%%%%%%%%%%%%%%%%%%%%%%%%%%%%%%%%%%%%%%%%%%%%%%%%%%
\acknowledgements
This work is supported in part by the RFBR grant 96-02-18897.
The autor thanks A. L. Kataev for useful disscussions and the
interest to this work, S. A. Larin for illuminating discussion
on perturbative corrections to the GLS sum rule
and W. Melnitchouk for useful comments and correspondence.

%%%%%%%%%%%%%%%%%%%%%%%%%%%%%%%%%%%%%%%%%%%%%%%%%%%%%%%%%%%%%%%%%%%%%%

%%%%%%%%%%%%%%%%%%%%%%%%%%%%%%%%%%%%%%%%%%%%%%%%%%%%%%%%%%%%%%%%%
%%TABLES

%
\begin{table}
\caption{Moments of the off-shell term $p^2\partial_{p^2}F_3^{\text{as}}(x)$
calculated at $p^2=M^2$ in the model described in Sec.\protect\ref{offshell}.
}
\label{tablo}
\begin{tabular}{c|ccccccccccc}
$n$ & 0 & 1 & 2 & 3 & 4 & 5 \\
\tableline
$p^2\partial_{p^2}M_n^{\text{as}}$ &
0 & $0.0394$ & $0.0165$ & $0.0069$ & $0.00308$ & $0.00146$ \\
\tableline
$n$ & 6 & 7 & 8 & 9 & 10\\
\tableline
$p^2\partial_{p^2}M_n^{\text{as}}$
& $7.32\cdot 10^{-4}$ & $3.82\cdot 10^{-4}$ & $2.04\cdot 10^{-4}$
& $1.11\cdot 10^{-4}$ & $6.09\cdot 10^{-5}$ \\
\end{tabular}
\end{table}
%

%%\end{document}

%% FIGURES

\vskip 2cm
\begin{figure}[htb]
\begin{center}
\input fig2.pic
\end{center}
\vskip 2cm
\caption{Nucleon Compton amplitude in the leading order in $Q^2$}
\label{fig2}
\end{figure}

\newpage

\begin{figure}[htb]
%\begin{center}
 \epsfbox{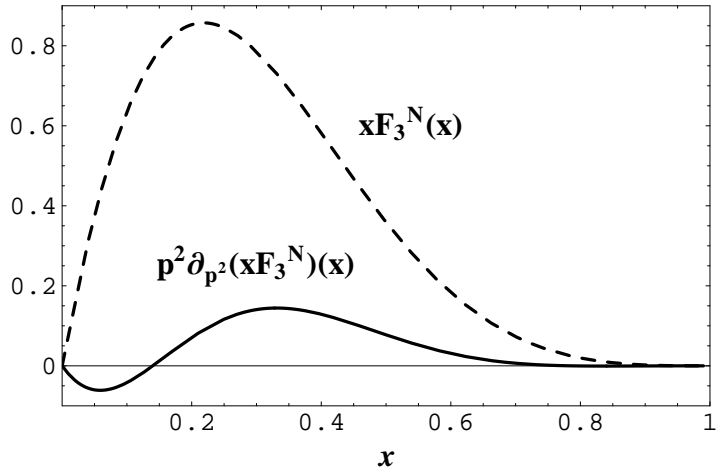}   %to be used with epsf.sty
\caption{The solid line shows the $x$ dependence of the derivative
$p^2\partial_{p^2}(xF_3)(x)$ calculated at the mass shell $p^2=M^2$
in the model described in Sec.\protect\ref{offshell}.
Also shown is the nucleon structure function $xF_3(x)$ (dashed line).
}
%\end{center}
\label{fig_f3}
\end{figure}
%
%\vskip 2cm

\newpage

\begin{figure}[htb]
 \epsfbox{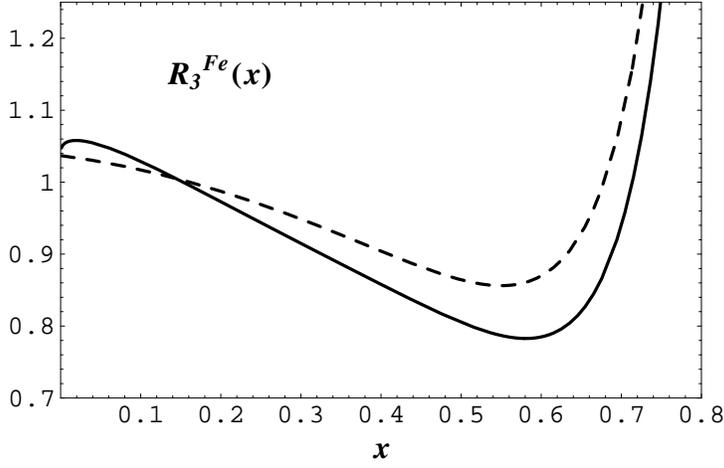}      %to be used with epsf.sty
        \epsfbox{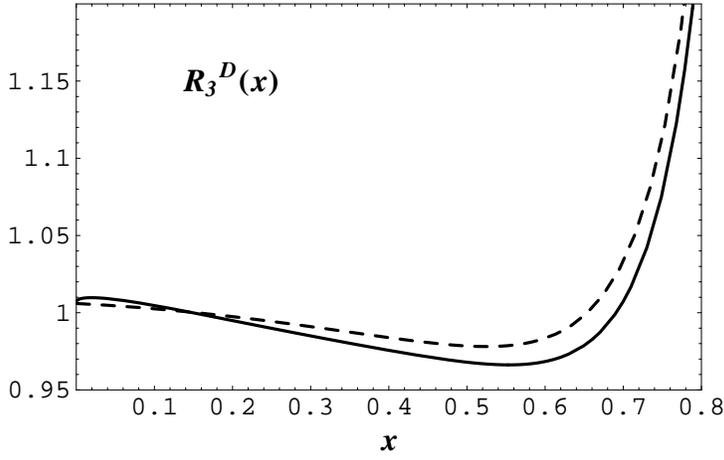}
\caption{The ratio $R_3(x,Q^2)=\frac1A F_3^A(x,Q^2)/F_3^N(x,Q^2)$ calculated
for the iron
(a)
and the deuteron (b)
nuclei by Eq.(\protect\ref{Dex}) at $Q^2=5\,$GeV$^2$.
The dashed line shows the results without off-shell corrections to
the structure function (i.e. without $\partial_{p^2}F_3$ term),
the solid line -- full calculation including the off-shell term.
}
\label{fig_ratio}
\end{figure}

\newpage

\begin{figure}[htb]
 \epsfbox{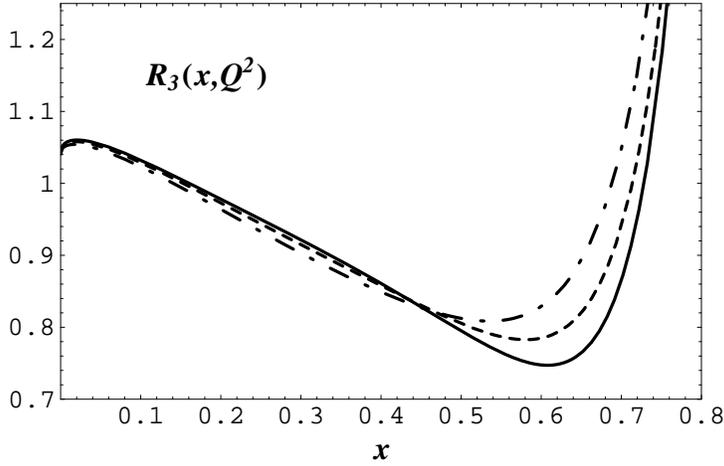}   %to be used with epsf.sty
\caption{Shown is the $Q^2$ dependence of the ratio
$R_3$ for the iron nucleus calculated by Eq.(\protect\ref{Dex}).
The dashed-dotted line corresponds to $Q^2=15\,$GeV$^2$,
the dashed line --- $Q^2=5\,$GeV$^2$,
while the solid line --- $Q^2=3\,$GeV$^2$.
}
\label{fig_ratio_q}
\end{figure}

\begin{figure}[htb]
 \epsfbox{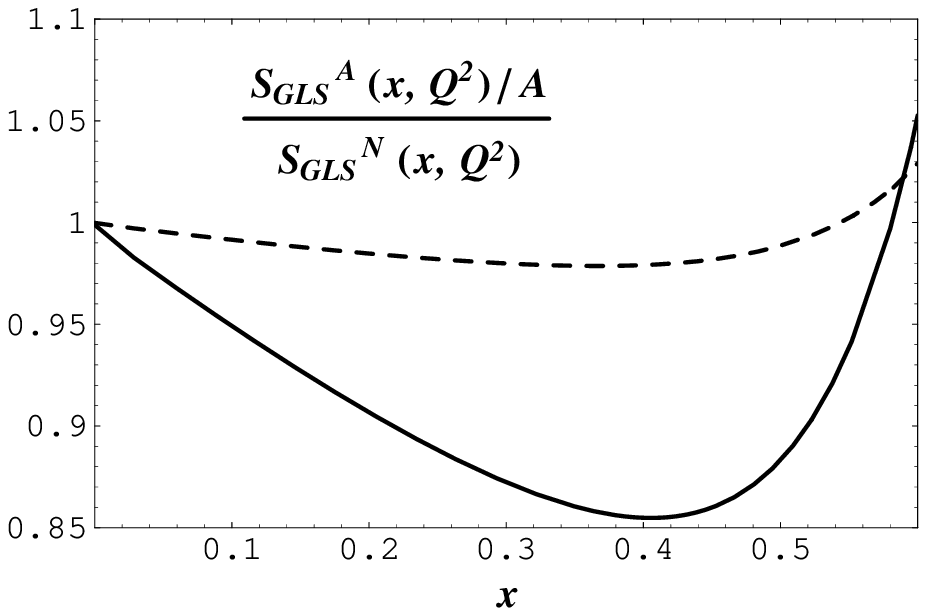}   %to be used with epsf.sty
\caption{Shown is the $x$ dependence of
the nucleus/nucleon ratio of the GLS integrals,
 $\frac1A S_{\text{GLS}}^A(x,Q^2)/S_{\text{GLS}}^N(x,Q^2)$,
calculated at $Q^2=5\,$GeV$^2$.
The dashed line corresponds to the deuteron,
while the solid line to the iron.
}
\label{fig_GLS_x}
\end{figure}


\begin{thebibliography}{99}


\bibitem{Arneodo94}
M. Arneodo, Phys. Rep. {\bf 240},  301  (1994).

\bibitem{GST95}
D. F. Geesaman, K. Saito, and A. W. Thomas,
Ann. Rev. Nucl. Part. Sci. {\bf 45}, 337 (1995).

\bibitem{CCFR:96}
{CCFR/NuTeV Collaboration, W. G. Seligman} {\it et~al.},
Phys. Rev. Lett. {\bf 79}, 1213 (1997).

\bibitem{CSB97}
J. M. Conrad, M. H. Shaevitz, and T. Bolton,
e-print hep-ex/9707015.

\bibitem{KM93}
B. Z. Kopeliovich, and P. Marage,
Int. J. Mod. Phys. {\bf A8}, 1513 (1993).

\bibitem{KKM95}
R. Kobayashi, S. Kumano, and M. Miyama,
Phys. Lett. {\bf B354}, 465 (1995).

\bibitem{SiTo95}
A. V. Sidorov, and M. V. Tokarev,
Phys. Lett. {\bf B358}, 353 (1995).

\bibitem{KKPS96}
A.~L. Kataev, A.~V. Kotikov, G. Parente, and A. Sidorov,
Phys. Lett. {\bf B388},  179  (1996);
Phys. Lett. {\bf B417}, 374 (1998).

%\bibitem{AKV85}
%S.~V. Akulinichev, S.~A. Kulagin, and G.~M. Vagradov,
%JETP Lett. {\bf 42} (1985) 127;
%Phys. Lett. {\bf B158},  485  (1985).

\bibitem{fm}
G. B. West, Ann. Phys. {\bf 74}, 464 (1972);\\
D. Kusno, and M. J. Moravcik, Phys. Rev. {\bf D20}, 2734 (1979);\\
A. Bodek, and J. L. Ritchie, Phys. Rev. {\bf D23}, 1070 (1981);\\
L. L. Frankfurt, and M. I. Strikman, Phys. Rep. {\bf 76}, 215 (1981).

\bibitem{binding}
S.~V. Akulinichev, S.~A. Kulagin, and G.~M. Vagradov,
JETP Lett. {\bf 42}, 127 (1985); Phys. Lett. {\bf B158},  485  (1985);
S.~V. Akulinichev, S. Shlomo, S.~A. Kulagin, and G.~M. Vagradov,
Phys. Rev. Lett. {\bf 55}, 2239 (1985);\\
B. L. Birbrair, A. B. Gridnev, M. B. Zhalov, E. M. Levin, and V. E. Starodubskii,
Phys. Lett. {\bf B166}, 119 (1986);\\
G. V. Dunne, and A. W. Thomas, Nucl. Phys. {\bf A455}, 701 (1986);\\
%%H. Jung, and G. A. Miller, Phys. Lett. {\bf B200}, 351 (1988);\\
C. Ciofi degli Atti, and S. Liuti, Phys. Lett. {\bf B225}, 215 (1989).

\bibitem{Ku89}
S.~A. Kulagin,
Nucl. Phys. {\bf A500},  653  (1989).

\bibitem{KPW94}
S.~A. Kulagin, G. Piller, and W. Weise,
Phys. Rev. {\bf C50},  1154  (1994).

\bibitem{KMPW95}
S.~A. Kulagin, W. Melnitchouk, G. Piller, and W. Weise,
Phys. Rev. {\bf C52},  932  (1995).

\bibitem{MeScTh94}
W. Melnitchouk, A.~W. Schreiber, and A.~W. Thomas,
Phys. Rev. {\bf C49}, 1183  (1994).

\bibitem{IoKhLi84}
B.~L. Ioffe, V.~A. Khoze, and L.~N. Lipatov,
{\em Hard processes:  Phenomenology, Quark-Parton Model}
(Elsevier Science Publishers, North  Holland, 1984).

\bibitem{mec}
M. Ericson and A.~W. Thomas, Phys. Lett. {\bf B128}, 112 (1983);\\
E. L. Berger, F. Coester, and R. B. Wirringa, Phys. Rev. {\bf D29}, 398 (1984);\\
%%B.~L. Friman, V.~R. Pandaripande, and R.~B. Wiringa, Phys. Rev. Lett. {\bf 51}, 763  (1983);\\
E. E. Saperstein, and M. Zh. Shmatikov, JETP Lett. {\bf 41}, 53 (1985);\\
L. P. Kaptari, A. I. Titov, E. L. Bratkovskaya, and A.Yu. Umnikov,
Nucl. Phys. {\bf A512}, 684 (1990).

\bibitem{EFP83}
R. K. Ellis, W. Furmanski, and R. Petronzio,
Nucl. Phys. {\bf B212}, 29 (1983).

\bibitem{Gur95}
S. A. Gurvitz, Phys. Rev. {\bf D52}, 1433 (1995).

\bibitem{fsi1}
E. Pace, G. Salme, and G. B. West, Phys. Lett. {\bf B273}, 205 (1991);\\
O. W. Greenberg, Phys. Rev. {\bf D47}, 331 (1993);\\
S. A. Gurvitz, and A. S. Rinat, Phys. Rev. {\bf C47}, 2901 (1993).

\bibitem{fsi2}
E. Pace, G. Salme, and F. M. Lev, Phys. Rev. {\bf C57}, 2655 (1998).

\bibitem{Kul_q90}
S. A. Kulagin, in Proc. Int. Seminar {\em Quarks '90},
edited by V. A. Matveev {\it et al.} (World Scientific,
Singapore, 1991).

\bibitem{STZ98}
M. Strikman, M. G. Tverskoy, and M. B. Zhalov, e-print nucl-th/9806099.

\bibitem{GLS69}
D. J. Gross, and C. H. Llewellyn-Smith,
Nucl. Phys. {\bf B14}, 337 (1969).

\bibitem{GLS_cor}
S. G. Gorishny, and S. A. Larin,
Phys. Lett. {\bf B172}, 109 (1986);
S. A. Larin, and J. A. M. Vermaseren,
Phys. Lett. {\bf B259}, 345 (1991).

\bibitem{BrKo87}
V. M. Braun, and A. V. Kolesnichenko,
Nucl. Phys. {\bf B283}, 723 (1987).

\bibitem{RR94}
G. G. Ross, and R. G. Roberts,
Phys. Lett. {\bf B322}, 425 (1994).

\bibitem{SFFB94}
I. Sick, S. Fantoni, A. Fabrocini, and O. Benhar,
Phys. Lett. {\bf B323}, 267 (1994).

\bibitem{NDM94}
D. Van Neck, A. E. L. Dieperink, and E. Moya de Guerra,
e-print nucl-th/9410019.


\end{thebibliography}
\end{document}